\documentclass[12pt, amsfonts, amssymb,color]{article}
\usepackage{amsmath,amssymb,amsfonts,latexsym}
\def\R{\mathbb{R}}
\def\K{\mathbb{K}}
\def\C{\mathbb{C}}
\begin{document}
%<<<<<<<<<<< enumeration of eqns section wise>>>>>>>>>>>>>>>>>>>

\renewcommand\theequation{\arabic{section}.\arabic{equation}}
\catcode`@=11 \@addtoreset{equation}{section}
%<<<<<<<<<<<<<<<<<<<<<<<<<<<<<<<<<>>>>>>>>>>>>>>>>>>>>>>>>>>>>>>>>>
\newtheorem{axiom}{Definition}[section]
\newtheorem{theo}{Theorem}[section]
\newtheorem{axiom2}{Example}[section]
\newtheorem{lem}{Lemma}[section]
\newtheorem{prop}{Proposition}[section]
\newtheorem{cor}{Corollary}[section]
\newcommand{\be}{\begin{equation}}
\newcommand{\ee}{\end{equation}}

\newcommand{\equal}{\!\!\!&=&\!\!\!}
\newcommand{\rd}{\partial}
\newcommand{\g}{\hat {\cal G}}
\newcommand{\bo}{\bigodot}
\newcommand{\res}{\mathop{\mbox{\rm res}}}
\newcommand{\diag}{\mathop{\mbox{\rm diag}}}
\newcommand{\Tr}{\mathop{\mbox{\rm Tr}}}
\newcommand{\const}{\mbox{\rm const.}\;}
\newcommand{\cA}{{\cal A}}
\newcommand{\bA}{{\bf A}}
\newcommand{\Abar}{{\bar{A}}}
\newcommand{\cAbar}{{\bar{\cA}}}
\newcommand{\bAbar}{{\bar{\bA}}}
\newcommand{\cB}{{\cal B}}
\newcommand{\bB}{{\bf B}}
\newcommand{\Bbar}{{\bar{B}}}
\newcommand{\cBbar}{{\bar{\cB}}}
\newcommand{\bBbar}{{\bar{\bB}}}
\newcommand{\bC}{{\bf C}}
\newcommand{\cbar}{{\bar{c}}}
\newcommand{\Cbar}{{\bar{C}}}
\newcommand{\Hbar}{{\bar{H}}}
\newcommand{\cL}{{\cal L}}
\newcommand{\bL}{{\bf L}}
\newcommand{\Lbar}{{\bar{L}}}
\newcommand{\cLbar}{{\bar{\cL}}}
\newcommand{\bLbar}{{\bar{\bL}}}
\newcommand{\cM}{{\cal M}}
\newcommand{\bM}{{\bf M}}
\newcommand{\Mbar}{{\bar{M}}}
\newcommand{\cMbar}{{\bar{\cM}}}
\newcommand{\bMbar}{{\bar{\bM}}}
\newcommand{\cP}{{\cal P}}
\newcommand{\cQ}{{\cal Q}}
\newcommand{\bU}{{\bf U}}
\newcommand{\bR}{{\bf R}}
\newcommand{\cW}{{\cal W}}
\newcommand{\bW}{{\bf W}}
\newcommand{\bZ}{{\bf Z}}
\newcommand{\Wbar}{{\bar{W}}}
\newcommand{\Xbar}{{\bar{X}}}
\newcommand{\cWbar}{{\bar{\cW}}}
\newcommand{\bWbar}{{\bar{\bW}}}
\newcommand{\abar}{{\bar{a}}}
\newcommand{\nbar}{{\bar{n}}}
\newcommand{\pbar}{{\bar{p}}}
\newcommand{\tbar}{{\bar{t}}}
\newcommand{\ubar}{{\bar{u}}}
\newcommand{\utilde}{\tilde{u}}
\newcommand{\vbar}{{\bar{v}}}
\newcommand{\wbar}{{\bar{w}}}
\newcommand{\phibar}{{\bar{\phi}}}
\newcommand{\Psibar}{{\bar{\Psi}}}
\newcommand{\bLambda}{{\bf \Lambda}}
\newcommand{\bDelta}{{\bf \Delta}}
\newcommand{\p}{\partial}
\newcommand{\om}{{\Omega \cal G}}
\newcommand{\ID}{{\mathbb{D}}}
\newtheorem{defi}{Definition}[section]
\newtheorem{exam}{Example}[section]
\def\R{\mathbb{R}}
\def\K{\mathbb{K}}
\def\C{\mathbb{C}}
\def\J{\mathbb{J}}

\title{ Nonlocal transformations of the Generalized Li\'{e}nard type
equations and dissipative Ermakov-Milne-Pinney systems\\}

\author{
Partha Guha\footnote{E-mail: partha@bose.res.in} \\
SN Bose National Centre for Basic Sciences \\
JD Block, Sector III, Salt Lake \\ Kolkata 700098,  India \\
\and
A Ghose-Choudhury\footnote{E-mail aghosechoudhury@gmail.com}
\\
Department of Physics \\ Diamond Harbour Women's University,\\
D.H. Road, Sarisha, West-Bengal 743368, India\\}

\date{ }

 \maketitle

\begin{abstract}
We employ the method of nonlocal generalized Sundman transformations to
formulate the linearization problem for equations of the generalized Li\'enard type and show that they may be
mapped to equations of the dissipative Ermakov-Milne-Pinney type. We obtain the
corresponding new first integrals of these derived equations, this method yields a natural generalization of the
construction of Ermakov-Lewis invariant for a time
dependent oscillator to (coupled) Li\'enard and Li\'enard type equations.
We also study the linearization problem for the coupled Li\'{e}nard equation
using  nonlocal transformations and derive coupled dissipative Ermakov-Milne-Pinney equation.
As an offshoot of this nonlocal transformation method when the standard Li\'{e}nard equation,
$\ddot{x}+f(x)\dot{x}+g(x)=0$,  is mapped to that of the linear harmonic oscillator equation we
obtain a relation between the functions $f(x)$ and $g(x)$
which is exactly similar to the condition derived in the context of isochronicity of
the Li\'{e}nard equation.

\end{abstract}

\bigskip

\paragraph{Mathematics Classification (2000) :} 34C14, 34C20.

\bigskip


\section{Introduction}
 The linearization of a nonlinear ordinary differential equation
 (ODE) has been an object of immense interest for many years. The
 most commonly employed method is to seek a point transformation
 such that the transformed ODE becomes linear and hence may be solved
 by some known method.\\
 Of late however, a number of attempts have
 been successfully made to tackle this problem by using
 Sundman transformations \cite{Sund}, which are
 nonlocal in character. Besides, second-order ODEs for which the
 linearization problem is well studied, Euler \textit{et al} \cite{EE,EESA,EWLE} have also extended
 their procedure to deal with higher-order (mainly third-order)
 ODEs.
The most general type of a nonlocal transformation (see for example, \cite{GCK} ) that may be
considered  is of the form \be\label{NLT1}dX=A(x,t) dx+B(x,t)
dt,\;\;\;\;dT=C(x,t)dx+D(x,t) dt.\ee The usual case of a point
transformation corresponds to the situation where
$A_t(x,t)=B_x(x,t)$ and $C_t(x,t)=D_x(x,t)$ so that
$$ \big(X(x,t),T(x,t)\big) \mapsto \big( F(x,t), G(x,t)\big), $$ such that $X(x,t) = F(x,t)$ and $T(x,t) = G(x,t)$.
In \cite{EE} as also in
\cite{Duarte1} it was assumed that $X(x,t)=F(x,t)$ but that
$C_t(x,t)\ne D_x(x,t)$, so that the temporal part is nonlocal. In
fact they took $C(x,t)=0$  so that $dT=D(x,t)dt$ assuming that
$D_x\ne 0$. Such a nonlocal transformation is commonly referred to
as a Sundman transformation \cite{BO}.

\bigskip

The Li\'{e}nard equation \cite{Lienard}
\be\label{Lie1}\ddot{x}+f(x)\dot{x}+g(x)=0,\ee has been
extensively studied owing to its diverse physical applications and
its appearance in the context of limit cycles of the Van der Pol
equation \cite{JS}.
The higher dimensional Li\'enard equation presents substantial additional difficulties which prevent
straightforward extensions of planar results. This is  particularly true for
 stability properties of equilibria, which are essential in studying the dynamics of perturbed systems.
In a  recent paper Briata and Sabatini \cite{BS} proved the asymptotic stability of the equilibrium solution of a class
of vector Li\'enard equations by means of the LaSalle invariance principle.

Recently Chandrasekar \textit{et al} \cite{Chan} have described a method for
the  linearization of coupled systems which may be briefly
summarized as follows. Given a system of coupled equations \be
\ddot{x} = \phi_1(x,y,\dot{x},\dot{y},t) \qquad \ddot{y} =
\phi_2(x,y,\dot{x},\dot{y},t), \ee one looks for a transformation
\be \omega_i = f_i(t,x,y), \qquad z_i = \int f_{i+2}(t,x.y)\, dt
\qquad i=1,2 \ee such that $\frac{d^2\omega_1}{dz_{1}^2} = 0$
and $\frac{d^2\omega_2}{dz_{2}^2} = 0$.
In the event that such a transformation exists   the coupled
system is said to be linearizable. The  procedure described by
them depends on the existence of two first integrals \be I_1 =
F(t,x,y,\dot{x},\dot{y})  \qquad I_2 = G(t,x,y,\dot{x},\dot{y}).
\ee  In fact they have shown that if the coupled system is
linearizable then the above transformation is completely
determined by these first integrals which must necessarily be of
the form $I_1 = \frac{1}{f_3}\frac{df_1}{dt}$ and $I_2 =
\frac{1}{f_4}\frac{df_2}{dt}$. This can be checked easily as
follows. Define $z_1$ and $z_2$ as $\frac{dz_1}{dt} = f_3$ and
$\frac{dz_2}{dt} = f_4$. This immediately yields
$$
I_1 = \frac{1}{\frac{dz_1}{dt}} \frac{df_1}{dt} = \frac{df_1}{dz_1} \qquad I_2 = \frac{1}{\frac{dz_2}{dt}} \frac{df_2}{dt} = \frac{df_2}{dz_2}.
$$
If we identify $\omega_1 \equiv f_1$ and   $\omega_2 \equiv f_2$ then $I_1 = \frac{d\omega_1}{dz_1}$ and
$$
\frac{d}{dz_1}\big(\frac{d\omega_1}{dz_1} \big) = \frac{dI_1}{dt}/\frac{dz_1}{dt} = 0.
$$
The major shortcoming of this method is that it requires explicit
knowledge of the first integrals  which in itself is a non trivial
problem.

\smallskip

A pioneering contribution towards the  linearization of the equations of motion occurring in celestial mechanics
was made by Sundman \cite{Sund} who introduced the transformation $dt = r d\tau$ in
his study of the $3$-body problem, where $r$ is the  dependent
variable (radial component). About a quarter of a century ago
Sundman's method was revitalized by Szebehely and Bond \cite{SzB},
who  considered a transformation of the dependent variable  $r
= F(\rho)$. The theoretical importance of the generalized
Sundman transformations stems from their occurrence in various areas
of mechanics and dynamical systems. In particular
transformations of  the Sundman type which are also referred to as non-point transformations
  by some authors \cite{CSL} are especially effective for obtaining
solutions of many nonlinear ODEs.

In \cite{Duarte1} the authors derived the most general condition
under which a second-order ordinary differential equation is
transformable to the linear equation
$X^{\prime\prime}(T)=0,$ (here $X^\prime=\frac{dX}{dT}$)
 under a generalized Sundman transformation.
In this communication we derive a systematic procedure to find the
first integral for SODE which are transformed to
$X^{\prime\prime}(T)+\omega^2X=0$ under generalized Sundman
transformation.

\bigskip

\noindent
{\bf Prelude, motivation and result :} 
This equation falls in the class of the
so-called projective connections and closely connected with different geometric problems, i.e., SODE of the form
\be \label{eq3rd}\ddot{x}+A_3(t;x)\dot{x}^3+A_2(t;x)\dot{x}^2+A_1(t;x)\dot{x}+A(t;x)=0.\ee
Lie was the first to study the linearization problem of SODE, he showed that every linearizable SODE can 
be recasted to the above form of equation (\ref{eq3rd}) and the coefficients satisfy the conditions
(see for example, \cite{Sookmee})
\be 3A_{3tt} - 2A_{2tx} + A_{1xx} = (3A_1A_3 - A_{2}^{2})_t - 3(AA_3)_x - 3A_3A_x +A_2A_{1x}, \ee
\be 3A_{xx} - 2A_{1tx} + A_{2tt} = 3(AA_3)_t + (A_{1}^{2} - 3AA_{2})_x + 3AA_{3t} - A_1A_{2t}. \ee
Such equations were studied by Lie \cite{Lie}, Tresse \cite{Tresse},
Cartan \cite{Cartan}, Liouville \cite{Liouville}, etc. (see, for example, \cite{Liouville}). 
The class of equations (\ref{eq3rd}) is closed under  generic point transformations.
It means that the transformed equation is again given by (\ref{eq3rd}) but with some other coefficients.
The problem of existence of the
change of variables that transforms equation (\ref{eq3rd}) into other with different coefficient is called the
Equivalence Problem. 

\smallskip

If an equation of this type admits an integral of the form $A(t, x)\dot{x} + B(t, x)$
then it must be of the form
$$ \ddot{x}+A_2(t;x)\dot{x}^2+A_1(t;x)\dot{x}+A_0(t;x)=0. $$
This result is contained in  \cite{Muriel}. A transparent motivation of this result in terms of
$\lambda$-symmetries is contained, for instance, in the paper \cite{Meleshko}, which is coauthored by the
authors the present paper. More motivations in terms of projective structures can be
also found in \cite{Bryant}. 

\smallskip

Another motivation for the present article stems from a recent paper of Padmanabhan \cite{Paddy}
in which he pointed out the physical basis for the Ermakov-Lewis invariant.
A generalization of Padmanabhan's original Lagrangian has been made by the authors of \cite{GR}
by including an additional potential term. Recently we reported the results of a further modification of the transformation used in \cite{Paddy} to derive generalizations of the time-dependent oscillator equation and its associated partner,
namely the Ermakov-Pinney equation. There is a hidden nonlocal transformation embedded in this
transformation. A few years ago we  examined the connection
between a time-dependent second-order ODE and the Ermakov-Pinney system \cite{GuGC},
where it was shown that by a {\it simple rational transformation of the dependent variable} one could easily
extract the well known Ermakov-Lewis invariant. In spite of such a large number of applications in physics
( for example, see \cite{Leach,RCS} for exhaustive references and historical background \cite{Leach}),
the Ermakov-Pinney equation in itself does not have any dissipation term, but the physical system demands a natural generalization of
the model by inclusion of the damping mechanism. It is known from \cite{Haas} that  damped
Ermakov-Pinney equations arise in quantum mechanical models
with dissipation.

\smallskip

It is clear from the work of Padmanabhan  \cite{Paddy,GuGC2} that there is a hidden nonlocal transformation embedded in
his construction and this provides the motivation to explore the connection between equations of the Li\'enard type
and the dissipative Ermakov-Pinney equations. In the second part of the paper we introduce the
damped Pinney equation considered in \cite{Haas} was defined as the model arising when a
damping term, linear in  the velocity, is included in the Pinney equation.
We generalize our results to coupled  Li\'enard equations and consider the mapping to a coupled
dissipative Ermakov-Milne-Pinney equation. As an offshoot of our program
we obtain the isochronous conditions stated by Sabatini  on $f(x)$ and $g(x)$ of the standard
Li\'enard equation \cite{Saba1,CD}.

\smallskip

The article is {\bf organized} as follows. In Section 2 we
introduce a nonlocal transformation and demonstrate the mapping of the the Li\'{e}nard type
equation to the Ermakov-Pinney type equation.  In Section 3 we study the mapping between
the generalized Li\'enard equation and generalized dissipative Ermakov-Milne-Pinney equation.
Section 4 is devoted to the coupled Li\'enard equation and the coupled dissipative Ermakov-Milne-Pinney equation.
In Section 5 we study the linearization of the standard Li\'enard equation, $\ddot{x} + f(x)\dot{x} + g(x) = 0$, 
using a nonlocal transformation and demonstrate that the isochronous conditions stated by Sabatini 
on $f(x)$ and $g(x)$ of the standard Li\'enard equation follows almost naturally.

\section{Linearization via Nonlocal transformations}

  In addressing the broader issue of
 linearizing a given second-order nonlinear ODE, it is useful if we enlarge the class of 
transformations usually considered to beyond point transformations and look for
 a more general class of transformations which are of a nonlocal character. The theory of nonlocal
 transformations can be traced back to the original works of
 Sundman (1912). Euler {\textit et al.} \cite{EE, EESA} have considered the class of Sundman
 transformations and have profitably used them to not only
 linearize  nonlinear ODEs but also to identify the so-called
 generalized Sundman symmetries of such equations. The present
 authors \cite{GCK} have further generalized the Sundman transformations and
 have applied them to deduce  first integrals of time-dependent
 second-order ODEs. Duarte \textit{et al.} \cite{Duarte1} have used nonlocal
 transformations for determining when a given ODE is equivalent to
 a linear differential equation. In the notation of \cite{Duarte1} it is usual to
 begin with a transformation of the form
 \be\label{Sun1}X=F(x,t),\;\;\;\;dT=G(x,t) dt,\ee and to determine the functions
 $F$ and $G$ such that the original second-order ODE
 $\ddot{x}=F(x,\dot{x})$ is mapped to a linear ODE, in
 particular to the free particle equation $d^2X/dT^2=0$.
Here we consider a modification of the above form and assume that
the nonlocal transformation is defined, in general, by
\be\label{NLT2}dh(X)=A(x,t)dx+B(x,t)dt,\ee where $h(X)$ is some
suitable function to be chosen while $G(x,t)=1$ so that $T=t$.\\

\subsection{Illustration: linearization of Li\'enard type equations}

As an illustration we consider a  second-order ordinary
differential equation (SODE) of the Li\'enard type having the form \be \label{s.1}
\ddot{x}+f(x)\dot{x}^2+ g(x)=0,\ee and
search for  a nonlocal transformation such that it is mapped to
the  simple harmonic equation
 \be \label{s.2}\ddot{X} + \omega^2 X = 0,\ee (here
$\dot{X}=\frac{dX}{dt}$) by the transformation \be \label{s.3}
\frac{dX}{X}=A(x,t)dx+B(x,t)dt.\ee It is a matter of straightforward computation to show that (\ref{s.1}) is mapped to
(\ref{s.2}) provided its coefficients satisfy the following
conditions: \begin{align}
\label{s.4} A^2+A_x - A(x)f(x) = 0,\\
\label{s.5} B_x+2A(x)B(x) = 0,\\
\label{s.6} A(x)g(x) - B^2 = \omega^2.
\end{align}

Suppose $A = v_x/v$, then after
solving the first and second equations stated above we get
$$
v_x = exp(\int^x f(s)ds), \qquad B(x) = \frac{1}{v^2}.
$$
The final equation then becomes
$$
\frac{v_x}{v}g(x) - \frac{1}{v^4} = \omega^2.
$$

Let us now set $ f(x) = \frac{\alpha}{x}$, which readily yields $v = \frac{x^{\alpha + 1}}{\alpha + 1}$.
Thus $A$ and $B$ are given by
\be
A(x) = \frac{\alpha + 1}{x}, \qquad B(x) = \frac{(\alpha + 1)^2}{x^{2(\alpha + 1)}}
\ee
and \be g(x) = \frac{(\alpha + 1)^3}{x^{4\alpha + 3}} + \omega^2 \frac{x}{(\alpha + 1)}. \ee
Therefore, the singular Sundman type transformation has the appearance
\be
X = x^{\alpha + 1}\exp\big({(\alpha + 1)^2}\int \frac{dt}{x^{2(\alpha + 1)}}\big) \qquad T = t,
\ee
and corresponds to the equation
\be
\label{s.n}
\ddot{x} + \frac{\alpha}{x}\dot{x}^{2} + \frac{1}{\alpha + 1}\omega^2 x +
\frac{(\alpha + 1)^3}{x^{4\alpha + 3}} = 0.
\ee
On the other hand by integrating (\ref{s.2}) we obtain
\be
(\frac{dX}{dT})^2 + \omega^2 X^2 = I(t,x,\dot{x}),
\ee
where $I(t,x,\dot{x})$ is the first integral. Using this recipe
the first integral of (\ref{s.n}) reads
$$
I = (\alpha + 1)^2 \big( x^{\alpha}\dot{x} + \frac{\alpha + 1}{x^{\alpha + 1}} \big)^2
+ \omega^2 \Big( x^{\alpha + 1}exp\big({(\alpha + 1)^2}\int \frac{dt}{x^{2(\alpha + 1)}}\big) \Big)^2.
$$

\smallskip

\paragraph{Remark} For $\alpha = 0$, equation (\ref{s.n}) boils down to the Ermakov-Pinney equation while
for $\alpha = -\frac{1}{2}$ it corresponds to a reduced version of an equation of the Gambier type, namely
\be
\ddot{x} - \frac{1}{2x}\dot{x}^{2} + 2\omega^2 x - \frac{1}{8x} = 0,
\ee
which incidentally has the additional feature of exhibiting the property of isochronicity.

\section{Nonlocal Sundman transformation of the generalized Li\'enard equation
and the generalized dissipative Ermakov-Milne-Pinney equations }

In this section we consider a general second-order differential equation (SODE) of
the type \be \label{gst.1}
\ddot{x}+A_2(t;x)\dot{x}^2+A_1(t;x)\dot{x}+A_0(t;x)=0\ee and look
for a generalized Sundman transformation such that it is mapped to
the following equation \be \label{gst.2}
X^{\prime\prime}(T)+\omega^2X=0.\ee 
Formally we define a generalized Sundman transformation for as follows.
 \begin{axiom}[Sundman transformation]
A coordinate transformation of the form
\be\label{gst.3} X(T)=F(t,x),\;\;\;dT=G(t,x)dt, \hskip 10pt
\frac{\partial F}{\partial x}\ne 0,\;\;\;G\ne 0\ee is said to be a
generalized Sundman transformation  if
differentiable functions $F$ and $G$ are determined such that
an $n$th-order ordinary differential equation
$$ x^{(n)}=w(t,x,\dot{x},\ddot{x},\dots,x^{(n-1)}), \qquad x^{(k)}={d^k x}/{dt^k}, $$
is transformed to the autonomous equation \be\label{2b}
X^{(n)}=w_0(X,X^\prime,\dots, X^{(n-1)}),\ee where
$X^\prime={dX}/{dT}$ etc.
 \end{axiom}

Straightforward computation then shows that  (\ref{gst.1})
is mapped to (\ref{gst.2}) provided its coefficients satisfy the
following conditions: \be\label{gst.4a}
\frac{F_{xx}}{F_x}-\frac{G_x}{G}=A_2(t,x)\ee

\be\label{gst.4b}
2\frac{F_{xt}}{F_x}-\frac{G_x}{G}\frac{F_t}{F_x}-\frac{G_t}{G}=A_1(t,x)\ee
\be\label{gst.4c}
\frac{F_{tt}}{F_x}-\frac{G_t}{G}\frac{F_t}{F_x}+\omega^2F\frac{G^2}{F_x}=A_0(t,x).\ee
Therefore given a SODE, so that the explicit form of the coefficients
$A_i(t,x)$'s are known, by solving the  set of equations
(\ref{gst.4a}) to (\ref{gst.4c}) if one can deduce the functions $F$
and $G$ then the linearizing transformation (\ref{gst.3}) may be obtained and consequently equation (\ref{gst.1}) may be  linearized to the equation of a linear  harmonic oscillator(\ref{gst.2}). \\
Integrating (\ref{gst.2}), we get \be \label{fi.1}
\left(\frac{dX}{dT}\right)^2+\omega^2X^2=
\left(\frac{F_x}{G}\dot{x}+\frac{F_t}{G}\right)^2+\omega^2F^2=I(t,x,\dot{x})=\mbox{constant},\ee
where $I(t,x,\dot{x})$ is the first integral. Having explained the
general idea behind construction of  the linearizing
transformation and a first integral for a given equation of the
type considered in (\ref{gst.1}), let us pass on to a description
of the the actual details of their construction.

\smallskip

Integrating (\ref{gst.4a}) w.r.t. $x$, we obtain \be \label{gst.5}
G=b(t,x)F_x,\ee where \be \label{gst.6}
b(t,x)=a(t)\;exp\left(-\int{A_2(t,x)dx}\right).\ee Here $a(t)$ is
an arbitrary function of $t$.
From equations (\ref{gst.4b}) and (\ref{gst.4c}), we have \be
\label{gst.7} S_x-\frac{b_x}{b}S=A_1(t,x)+\frac{b_t}{b},\ee \be
\label{gst.8} S_t-\frac{b_t}{b}S+\omega^2b^2FF_x=A_0(t,x),\ee
where \be \label{gst.9}S=\frac{F_t}{F_x}.\ee
Solving equation (\ref{gst.7}), we find that \be \label{gst.10}
S=c(t)b(x,t)+b(x,t)\int{\frac{A_1(t,x)+\frac{b_t}{b}}{b(x,t)}dx},\ee
where $c(t)$ is an arbitrary function of $t$. The explicit form
of $F$ can now be determined by substituting the expression for $S$
into (\ref{gst.9}) and solving the resultant first-order partial
differential equation for $F$ namely, \be F_t-SF_x=0.\ee Once $F$
is known $G$ can be found from  (\ref{gst.5}) and (\ref{gst.6})
which in turn provide us the GST as given in
(\ref{gst.3}).\\
Note that when the expressions for $S$ and $F$ are substituted into  (\ref{gst.8})  then the latter  must be identically
 satisfied.\\

We illustrate the procedure described above with a few simple
but nontrivial examples. All these examples are related to the parametric extensions
of the Gambier equation. 
\be\label{G2}
\ddot{x}=\left(1-\frac{1}{n}\right)
\frac{\dot{x}^2}{x}+a\frac{n+2}{n} x\dot{x} +b\dot{x}
-\left(1-\frac{2}{n}\right) s\frac{\dot{x}}{x}
-\frac{a^2}{n}x^3+(\dot{a}-ab)x^2+\left(cn-\frac{2as}{n}\right)x-bs
-\frac{s^2}{nx}.\ee 

In our illustration we assume all the coefficients are functions of the independent variable $t$. 

\smallskip

\begin{prop}\label{prop.1}
A time dependent first integral of the second order equation of
the form
$$\ddot{x}+\frac{\alpha}{x}\dot{x}^2+\frac{\beta}{t}\dot{x}+A_0(x,t)=0$$
is given by the function
$$I(t,x,\dot{x})=
\left(\frac{x^\alpha}{a}\dot{x}+ \frac{x^{\alpha+1}(\beta
a+t\dot{a})}{t(\alpha+1)a^2}\right)^2+\omega^2\left(x^{\alpha+1}t^\beta
a\right)^2,$$ where
$$A_0(x,t)=\left(\frac{\ddot{a}}{a}-2\frac{\dot{a}^2}{a^2}+
\frac{\dot{\beta}}{t}-\frac{\beta}{t^2}-\frac{\dot{a}\beta}{a
t}\right)\frac{x}{\alpha+1}+\omega^2\lambda
(\alpha+1)a^{2\lambda+2}t^{2\lambda \beta}x^{2\lambda
\alpha+2\lambda-2\alpha-1}.$$  $\alpha, \beta$  are constants,
$\lambda$ is an integer and $a(t)$ is an arbitrary function of
$t$.
\end{prop}
\smallskip
%**********************************************************
\noindent {\bf Proof}: In the above proposition
$A_2(x,t)=\frac{\alpha}{x}$ and $A_1(x,t)=\frac{\beta}{t}$. Our
main aim is to find $F$ and $G$.  From (\ref{gst.6}) we find that \be
\label{p1.1} b(x,t)=\frac{a}{x^\alpha}.\ee Substituting $A_1(x,t)$
and $b(x,t)$ in (\ref{gst.10}), we have  \be \label{p1.3}
S(x,t)=\left(\frac{\dot{a}}{a}+\frac{\beta}{t}\right)\frac{x}{\alpha+1},\ee
where we have set $c(t)=0$. Now, substituting $S$ in (\ref{gst.9})
we have the first order partial differential equation \be
\label{p1.4}\frac{F_t}{\frac{\dot{a}}{a}+\frac{\beta}{t}}-\frac{F_x}{\frac{x}{\alpha+1}}=0.\ee
By using the method of characteristics we obtain the general of
$F(x,t)$ in the form \be \label{p1.5} F(x,t)=J(a\;
x^{\alpha+1}t^\beta),\ee where $J(\xi)$ is any arbitrary function
of the characteristic coordinate $\xi=a\; x^{\alpha+1}t^\beta$.
Assuming $F=(a\; x^{\alpha+1}t^\beta)^\lambda$, we have from
(\ref{gst.5}) the following expression for $G$, \be \label{p1.6}
G=\lambda (\alpha+1)a^{\lambda+1}t^{\beta
\lambda}x^{(\alpha+1)(\lambda-1)}.\ee Therefore, the GST is given
by,
$$X=a\; x^{\alpha+1}t^\beta)^\lambda,$$
$$dT=\lambda
(\alpha+1)a^{\lambda+1}t^{\beta
\lambda}x^{(\alpha+1)(\lambda-1)}dt.$$

Note that the expressions for $S$ and $F$ as given in (\ref{p1.3})
and (\ref{p1.5}) respectively when substituted in (\ref{gst.8})
indeed give $A_0(x,t)$. The
expression for the first integral is obtained from
(\ref{fi.1}) after substituting $F$ and $G$ from (\ref{p1.5}) and
(\ref{p1.6}) respectively.
%************************************************************
\begin{prop}\label{prop.2}
A time dependent first integral of a second-order equation of
the form
\be\label{GEq} \ddot{x}+\frac{\alpha}{x}\dot{x}^2+\beta(t)x^\lambda\dot{x}+A_0(x,t)=0\ee is given by the
function
$$I(t,x,\dot{x})=
\left(\frac{x^\alpha}{a}\dot{x}+\frac{\beta(t)x^{\lambda+
\alpha+1}}{\lambda+\alpha+1}\right)^2+\omega^2\left(\beta_1(t)-\frac{\lambda+\alpha+1}
{\lambda}x^{-\lambda}\right)^2$$ where
$$A_0(x,t)=\frac{\dot{\beta}(t)x^{\lambda+1}}{\lambda+\alpha+1}+\omega^2(\lambda+\alpha+1)
\beta_1(t)x^{-2\alpha-\lambda-1}-\omega^2(\lambda+\alpha+1)^2\frac{x^{-2\alpha-2
\lambda-1}}{\lambda},$$
$$\beta_1(t)=\int{\beta(t)dt}.$$
\end{prop}
\smallskip
%**********************************************************
\noindent {\bf Proof:} \,\, In the above proposition
$A_2(x,t)=\frac{\alpha}{x}$ and $A_1(x,t)=\beta(t)x^\lambda$.
Once again our main aim is to find $F$ and $G$
under the transformation $X = F(t; x)$ and $dT = G(t; x)dt$.
For this, first of all we
evaluate $b(x,t)$. From (\ref{gst.6}) we have \be \label{p2.1}
b(x,t)=\frac{a}{x^\alpha}.\ee Substituting $b(x,t)$ in
(\ref{gst.7}), we have \be \label{p2.2}
S_x+\frac{\alpha}{x}S=\frac{\dot{a}}{a}+\beta(t)x^\lambda.\ee A
particular solution of (\ref{p2.2}) is clearly given by \be
\label{p2.3}
S(x,t)=\frac{\dot{a}}{a}x+\beta(t)\frac{x^{\lambda+\alpha+1}}{\lambda+\alpha+1},\ee
where we have set the constant of integration to be zero. Setting
the arbitrary function $a(t)=\mbox{constant}=1$, we have \be
\label{p2.4}
S=\beta(t)\frac{x^{\lambda+\alpha+1}}{\lambda+\alpha+1}.\ee Now,
substituting $S$ in (\ref{gst.9}) we have the first order partial
differential equation
$$\frac{x^{\lambda+\alpha+1}}{\lambda+\alpha+1}F_x-\frac{F_t}{\beta(t)}.$$
By using the method of characteristics we obtain the general of
$F(x,t)$ in the form \be \label{p2.5}
F=J(\beta_1(t)-\frac{\lambda+\alpha+1}{\lambda}x^{-\lambda}),\ee
where $J(\xi)$ is any arbitrary function of the characteristic
coordinate
$\xi=\beta_1(t)-\frac{\lambda+\alpha+1}{\lambda}x^{-\lambda}$.
Setting
$F=\beta_1(t)-\frac{\lambda+\alpha+1}{\lambda}x^{-\lambda}$, we
have from (\ref{gst.5}) and using (\ref{gst.6}),
$G=\frac{\lambda+\alpha+1}{x^{\lambda+\alpha+1}}$. Therefore, the
GST looks like,
$$X=\beta_1(t)-\frac{\lambda+\alpha+1}{\lambda}x^{-\lambda},$$
$$dT=\frac{\lambda+\alpha+1}{x^{\lambda+\alpha+1}}dt.$$
It is easy to verify that these expression for $S$ and $F$ gives
the required expression for $A_0(x,t)$ when substituted in
(\ref{gst.8}). Again, we find the required expression for the
first integral after substituting $F$ and $G$ in (\ref{fi.1}).
$\boxed{}$

\bigskip

\begin{cor}
For $\alpha = -3$, $\lambda = 4$, $\beta(t) = 1/2$ equation (\ref{GEq}) reduces to
$$ \ddot{x}-\frac{3}{x}\dot{x}^2+ \frac{1}{2}x^4\dot{x} + \omega^2 x = \frac{\omega^2}{x^3}, $$
where `` derivative free'' terms coincide with the Ermakov-Pinney equation.
\end{cor}

\smallskip

%*************************************************************
\begin{prop}
A time dependent first integral of the second order equation of
the form
$$\ddot{x}+3\alpha(t)x\dot{x}+\frac{3}{2}\dot{\alpha}(t)x^2
+\frac{2\omega^2}{3x^2}\alpha_1(t)-\frac{4\omega^2}{9x^3}=0$$ is
given by the function
$$I=(t,x,\dot{x})=\left(\dot{x}+\frac{3\alpha(t)x^2}{2}\right)^2+
\omega^2\left(\alpha_1(t)-\frac{2}{3x}\right)^2,$$ where
$\alpha_1(t)=\int{\alpha(t)dt}$.
\end{prop}

\smallskip
%**********************************************************
\noindent {\bf Proof}: In the above proposition $$ A_2(x,t)=0,\qquad
A_1(x,t)=3\alpha(t)x, \,\,\,\,\, \hbox{ and }\,\,\,\,\,\,\, A_0(x,t)=\frac{3}{2}\dot{\alpha}(t)x^2
+\frac{2\omega^2}{3x^2}\alpha_1(t)-\frac{4\omega^2}{9x^3}. $$

Our main aim is to find $F$ and $G$. For this, first of all we
evaluate $b(x,t)$. From (\ref{gst.6}) we have \be \label{p3.1}
b(x,t)=a(t).\ee Again, from (\ref{gst.5}), $G$ can be written as
\be \label{p3.2} G=a(t)F_x.\ee Therefore, the equations
(\ref{gst.7}) and (\ref{gst.8}) can be written as  \be
\label{p3.3} S_x=3\alpha(t)x\;\dot{x}+\frac{\dot{a}}{a},\ee \be
\label{p3.4}
S_t-\frac{\dot{a}}{a}S+\omega^2a^2FF_x=\frac{3}{2}\dot{\alpha}(t)x^2
+\frac{2\omega^2}{3x^2}\alpha_1(t)-\frac{4\omega^2}{9x^3}.\ee

A particular solution of (\ref{p3.3}) is clearly given by \be
\label{p3.5}S=\frac{3}{2}\alpha(t)x^2+\frac{\dot{a}}{a}x,\ee where
we have set the constant of integration to be zero. Setting the
arbitrary function $a(t)=\mbox{constant}=1$, we have \be
\label{p3.6} S=\frac{3}{2}\alpha(t)x^2.\ee Now, substituting $S$
in (\ref{gst.9}) we have the first order partial differential
equation
$$\frac{3}{2}x^2F_x-\frac{1}{\alpha(t)}F_t=0$$
By using the method of characteristics we obtain the general of
$F(x,t)$ in the form \be \label{p3.6}
F=J(\alpha_1(t)-\frac{2}{3x}),\ee where $J(\xi)$ is any arbitrary
function of the characteristic coordinate
$\xi=\alpha_1(t)-\frac{2}{3x}$. Setting
$F=\alpha_1(t)-\frac{2}{3x}$, we have from (\ref{p3.2}),
$G=\frac{2}{3x^2}$. Therefore, the GST looks like,
$$X=\alpha_1(t)-\frac{2}{3x},$$
$$dT=\frac{2}{3x^2}dt.$$
It is easy to verify that the equation (\ref{p3.4}) is identically
satisfied for these $S$ and $F$. Again, we find the required
expression for the first integral after substituting $F$ and $G$
in (\ref{fi.1}).
%****************************************************************
\begin{prop}
A time dependent first integral of the second order equation of
the form
\be\label{GEP}\ddot{x}+\frac{\alpha}{x}\dot{x}^2+\left(\frac{\ddot{a}}{a}-2\frac{\dot{a}^2}{a^2}
\right)\frac{x}{\alpha+1}+\lambda
\omega^2(\alpha+1)a^{2\lambda+1}x^{2\lambda\alpha+2\lambda-2\alpha-1}=0 \ee
 is given by the
function
$$I(t,x,\dot{x})=
\left(\frac{x^\alpha}{a}\dot{x}+\frac{\dot{a}x^{\alpha+1}}{a^2(\alpha+1)}\right)^2+
\omega^2(a x^{\alpha+1})^{2\lambda}$$ where $\alpha, \lambda$ is
constant and $a(t)$ is an arbitrary function of $t$.
\end{prop}

\smallskip
%**********************************************************
\noindent {\bf Proof}: In the above proposition $A_2(x,t)=0$,
$A_1(x,t)=\frac{\alpha}{x}$ and
$$ A_0(x,t)=\left(\frac{\ddot{a}}{a}-2\frac{\dot{a}^2}{a^2}
\right)\frac{x}{\alpha+1}+\lambda
\omega^2(\alpha+1)a^{2\lambda+2}x^{2\lambda\alpha+2\lambda-2\alpha-1}.$$
The proof is similar to proposition (\ref{prop.1}) since this
proposition is a particular case of Proposition (\ref{prop.1})
with $\beta(t)=0$.
$\Box$

\bigskip

\noindent
It should be noted that equation (\ref{GEP}) is the master equation of many Ermakov-Pinney type equation.
We give few of them.

\begin{cor}
(a) If we set $\alpha = 0$ and $\lambda = -1$, then the equation (\ref{GEP}) reduces  to
an equation of the Ermakov-Pinney type, \textit{viz}
\be \ddot{x} + \left(\frac{\ddot{a}}{a}-2\frac{\dot{a}^2}{a^2}\right)x = \frac{\omega^2}{ax^3}, \ee
and the corresponding first integral is given by
$$I_{gEP}(t,x,\dot{x})=
\left(\frac{\dot{x}}{a} + \frac{\dot{a}x}{a^2}\right)^2 +
\omega^2(a x)^{-2}.$$
(b) For  $\alpha = - \frac{3}{2}$, $\lambda = -1$, we obtain  the (parametric) Kummer-Schwarz equation, \textit{viz}
\be \ddot{x} - \frac{3}{2}\frac{\dot{x}^{2}}{x} -2b(t)x + \frac{\omega^2}{2a}x^3 = 0, \ee
where $b(t) = (\frac{\ddot{a}}{a}-2\frac{\dot{a}^2}{a^2})$.
\end{cor}

\smallskip

\begin{cor}
If we set  $\alpha = -\frac{1}{2}$, $a =1$, $\lambda = 1$ and shifting $\omega^2 \mapsto 2\omega^2$,
then equation (\ref{GEP}) reduces  to
\be
\ddot{x} - \frac{\dot{x}^{2}}{x} + \omega^2 x = 0,
\ee
whose first integral is given by $I = \frac{\dot{x}^{2}}{x} + 2\omega^2 x$.
\end{cor}

\section{ Mapping of the coupled Li\'enard equation to a coupled dissipative
Ermakov-Milne-Pinney equation via nonlocal
transformation}

We consider the following coupled Li\'enard equation
\be
\ddot{x} + f_1(x,y)\dot{x} + g_1(x,y) = 0, \qquad \ddot{y} + f_2(x,y)\dot{y} + g_2(x,y) = 0,
\ee
where $f$ is defined on an open connected subset $\Omega$ of ${\Bbb R}^2$. Briata and Sabatini \cite{BS} studied
a coupled equation where the coupling is entirely due to the dissipative terms and $g({\bf x}) = \big(g_1(x), g_2(y)\big)$.

\begin{prop}
The coupled Li\'enard equation $\ddot{x} + f_1(x,y)\dot{x} + g_1(x,y) = 0$ and $\ddot{y} + f_2(x,y)\dot{y} + g_2(x,y) = 0$
is mapped to the equations of two linear harmonic oscillator $\ddot{X} + \omega^2 X = 0$ and  $\ddot{Y} + \omega^2 Y = 0$,
under the following nonlocal transformation
\be
\frac{dX}{X} = A_1(x,y)dx + B_1(x,y)dy + C_1(x,y)dt, \qquad T = t
\ee
\be
\frac{dY}{Y} = A_2(x,y)dx + B_2(x,y)dy + C_2(x,y)dt, \qquad T = t,
\ee
where $$A_i = B_i = \frac{1}{x+y},\,\,\,\, (x+y)^2C_1(x,y) = \int_{0}^{x}(s+y)f_1(s,y)ds, \,\,\,\,
(x+y)^2C_2(x,y) = \int_{0}^{y}(x+s)f_2(x,s)ds $$ provided
$$
g_1 + g_2 = \omega^2(x+y) + \frac{1}{(x+y)^3} \Big( \int_{0}^{x}(s+y)f_1(s,y)ds \Big)
= \omega^2(x+y) + \frac{1}{(x+y)^3} \Big(\int_{0}^{y}(x+s)f_2(x,s)ds \Big).
$$
The consistency condition implies $(f_1 - f_2) = (x+y)(f_{2x} - f_{1y})$.

\end{prop}

{\bf Proof :} From ${dX}/{X} = A_1(x,y)dx + B_1(x,y)dy + C_1(x,y)dt$ we have
$$
\frac{\dot{X}}{X} = A_1(x,y)\dot{x} + B_1(x,y)\dot{y} + C_1(x,y),
$$
which after further differentiation we obtain
$$
\ddot{X}= \big[A_{1x}\dot{x}^{2} + (A_{1y} + B_{1x})\dot{x}\dot{y} + B_{1y}\dot{y}^{2} + (C_{1x} - A_1f_1)\dot{x}
+ (C_{1y} - B_1f_2)\dot{x} - A_1g_1 - B_1g_2 \big]X
$$
$$
+ \big[ A_{1}\dot{x} +  B_{1}\dot{y} + C_1 \big]^2 X.
$$

Now we set
$$
A_{1}^{2} + A_{1x} = 0
$$
$$
B_{1x} + 2A_1B_1 + A_{1y} = 0
$$
$$
C_{1x} + 2A_1C_1 - A_1f_1 = 0
$$
$$
C_{1y} + 2B_1C_1 - B_1f_2 = 0
$$
$$
\omega^2 = A_1g_1 + B_1g_2 - C_{1}^{2}.
$$
A particular solution of the first two equations is obvious, given by
$ A_1 = 1/ x+y = B_1$. Here we have chosen the constant of integration to be zero.
Inserting these values of $A_1$ and $B_1$ we obtain the expression of $C_1$.
The last equation yields the value of $g_1+g_2$.
From the consistency condition of
$$
\frac{\partial}{\partial x}\big( (x+y)^2 C_1 \big) = (x+y)f_1(x,y) \qquad \frac{\partial}{\partial y}\big( (x+y)^2 C_1 \big)= (x+y)f_2(x,y)
$$
we obtain $(f_1 - f_2) = (x+y)(f_{2x} - f_{1y})$.
$\Box$

\smallskip

Thus a sufficient condition is proposed for the
simultaneous linearization of the coupled equations.
Once again we assume our transformation is defined everywhere except at the singular points.

\bigskip

{\underline{\bf Illustration: }}\\
Let us take $$ f_1(x,y) = \frac{1}{x^3y}\qquad \hbox{ and } \qquad f_2(x,y) = \frac{1}{xy^3}. $$
This immediately yields $C_1 = - 1 /2x^2y^2$ and
$$ g_1 + g_2 = \big(\omega^2 + \frac{1}{4x^4y^4} \big)(x+y). $$ The natural choice of $g_1$ and $g_2$ are
$$
 g_1(x,y) = \omega^2 x +  \frac{1}{4x^3y^4} \qquad  g_2(x,y) = \omega^2 y +  \frac{1}{4x^4y^3}.
$$
Thus we obtain a coupled version
$$
\ddot{x} + \frac{\dot{x}}{x^3y} + \omega^2x + \frac{1}{4x^3y^4} = 0
$$
$$
\ddot{y} + \frac{\dot{y}}{xy^3} + \omega^2y + \frac{1}{4x^4y^3} = 0
$$
of the Ermakov-Milne-Pinney equation. The first integral of this coupled equation is
given by
\be
I = (\dot{x}y - x\dot{y}) - \frac{1}{2}(\frac{1}{x^2} - \frac{1}{y^2}).
\ee
\section{Linearization and isochronous conditions for the Li\'enard equation}
The Li\'{e}nard equation $\ddot{x}+f(x)\dot{x}+g(x)=0$ has been
studied extensively owing to its applications and associated
mathematical properties such as the existence of limit cycles for
suitable choices of the functions $f(x)$ and $g(x)$. Furthermore
this equation also displays isochronous behavior when $f(x)$ and
$g(x)$ bear a specific relationship together with certain
conditions on their character as will be stated in the sequel. \\

However, we will first  focus on its linearization via a nonlocal
transformation, which is defined everywhere except at the singular points,
to the equation of a linear harmonic oscillator.

\bigskip

  \begin{prop}\label{p1} The Li\'{e}nard equation $\ddot{x}+f(x)\dot{x}+g(x)=0$ is
   mapped to  the equation of a linear harmonic oscillator,
   $\ddot{X}+\omega^2X=0$, under the following nonlocal transformation
   $$\frac{dX}{X}=A(x)dx+B(x)dt,\;\;\;\;T=t$$ where $A(x)=1/x$ and
   $x^2B(x)=\int_0^x sf(s)ds$ provided $$g(x)=\omega^2
   x+\frac{1}{x^3}\left(\int_0^x sf(s)ds\right)^2.$$
   \end{prop}
\textbf{Proof:} From $dX/X=A(x)dx+B(x)dt$ we have
$$\frac{\dot{X}}{X}=A(x)\dot{x}+B(x)$$ which implies upon further
differentiation (note that  since $T=t$ we continue   denoting the
derivatives with overdots)
$$\ddot{X}=\left[(A_x+A^2)\dot{x}^2+(B_x+2A(x)B(x)-A(x)f(x))\dot{x}+(B^2(x)-A(x)g(x))\right]X.$$
Next we set
\begin{align}
A_x+A^2&=0\\
B_x+2A(x)B(x)-A(x)f(x)&=0\\
B^2(x)-A(x)g(x)&=-\omega^2.\end{align} A particular solution of
the first of these equations is obviously $A(x)=1/x$ where we have
chosen the constant of integration (which can be an arbitrary
function of $t$ ) to be zero. Inserting this expression for $A(x)$
into the second equation we get after integration
$$B(x)=\frac{1}{x^2}\int_0^xs f(s) ds.$$ Finally it follows from
the last equation that $g(x)=(B^2+\omega^2)/A(x) =
x(B^2+\omega^2)$, since $A(x)= 1/x$, so that
\be\label{maineqn}g(x)=\omega^2 x+\frac{1}{x^3}\left(\int_0^x s
f(s) ds\right)^2.\ee
$\Box$

\bigskip

{\bf Remark about isochronicity condition :}
Here we dwell on the aspect of isochronicity of such an
equation. In \cite{Saba1} the author has studied the monotonicity
properties of the period function of (\ref{Lie1}). In particular
it is shown that if the functions $f$ and $g$ be analytic, $g$
odd, $f(0)=g(0)=0$ and $g^\prime(0)>0$ then the origin is an
isochronous center if and only if $f$ is odd and
\be\label{tau}\tau(x):=\left(\int_0^x s f(s) ds\right)^2-x^3
(g(x)-g^\prime(0)x)\equiv 0.\ee
Christopher and Devlin \cite{CD} is seemingly re-proved by using a
different technique.

In all our subsequent calculations we will
take $\omega=1$ without loss of generality. This is in fact
identical
to the condition derived by Sabatini \cite{Saba1} given by (\ref{tau}).\\
 It is evident that here the function $h(X)=\log{X}$ while
\be\label{NLT3}X=F(x,t)=x\exp({\int B(x) dt}),\;\;\;T=t.\ee
It is clear that when $\omega=0$, then we can linearize the
Li\'{e}nard equation to $\ddot{X}=0$, by means of the above
nonlocal transformation provided
$$g(x)=\frac{1}{x^3}\left(\int_0^x s f(s)
ds\right)^2.$$
\section{Conclusion}
We have studied several nonlinear differential equations of Generalized Li\'enard type using nonlocal transformations. 
In particular, we have specifically chosen to map the initial nonlinear equation to that of a linear harmonic oscillator 
under a nonlocal/Sundam transformation and have even extended the procedure to a system of coupled nonlinear ODEs. 
As illustrated by our examples they can be used profitably to obtain the first integrals of several nonlinear 
ODEs displaying different types of damping /velocity dependance, and nature of these equations are  
damped Ermakov-Pinney type.  Thus we have generalized the construction of Ermakov-Lewis invariant for a time
dependent oscillator to (coupled) Li\'enard and Li\'enard type equations, in fact, it was one of our motivation to
write this paper.   
As the preceding section illustrates nonlocal transformations 
are quite useful for deriving the conditions to be obeyed by the functions $f$ and $g$ of the standard Li\'{e}nard equation 
to allow for isochronous motions. This utility can be extended to a much wider context, by treating such transformations as legitimate entities.

\section*{Acknowledgements}

We are profoundly grateful to Dr. Barun Khanra for  enlightening discussions and collaboration at
the primary stages of the work.
We wish to thank  Basil Grammaticos and Peter
Leach for valuable suggestions and constant encouragements. We also wish to thank Thanu Padmanabhan and Haret Rosu
for discussions over emails and
encouragements. We are indebted to an anonymous reviewer for providing insightful comments and
suggestions.


\begin{thebibliography}{99}

\bibitem{BO} Berkovich, L.M. and Orlova, I.S., {\em The exact linearization of some classes
of ordinary differential equations for order $n > 2$}. Proceedings
of Institute of Mathematics of NAS of Ukraine 30, part 1, 90-98.

\bibitem{BS} F. Briata and M. Sabatini, {\em Asymptotic stability and periodic solutions of vector Li\'enard equations}, arXiv:1008.3118v1 [math.CA].

\bibitem{Bryant} R. L. Bryant, G. Manno, V. S. Matveev, A solution of a problem of Sophus Lie:
normal forms of two-dimensional metrics admitting two projective vector fields,
Math. Ann. 340 (2008), no. 2, 437–463.

\bibitem{Cartan} E. Cartan, {\em Sur les vari\'et\'es \'a connexion projective}, Bull. Soc. Math. France 
52 (1924) 205-41.

\bibitem{Chan} V.K. Chandrasekar, M. Senthilvelan and M.L. Lakshmanan, {\em Unusual Li\'enard-type nonlinear
oscillator}. Phys. Rev. E 72 (2005) 066203.

\bibitem{CSL} Chandrasekar, V. K.; Senthilvelan, M.; Lakshmanan, M. {\em A unification in the theory of
linearization of second-order nonlinear ordinary differential
equations}.  J. Phys. A  39 (2006),  no. 3, L69--L76.

\bibitem{CD} C. Christopher and J. Devlin, {\em On the classification of Li\'{e}nard systems with
amplitude-independent periods},  J. Differential Equations  200
(2004),  no. 1, 1–17.

\bibitem{RCS} R. Campoamor-Stursberg, {\em Deformations of Lagrangian systems preserving a
fixed subalgebra of Noether symmetries}, preprint, $https://web.ma.utexas.edu/mp_arc/c/15/15-85.pdf$.
Acta Mech. 229 (2018), no. 1, 211-229.

\bibitem{Duarte1} L.G.S. Duarte, Moreira, I. C. and Santos, F. C. {\em Linearization under non-point
transformations}. J.Phys. A 27 (1994) L739-L743.

\bibitem{Duarte} L.G.S. Duarte, S. E. S. Duarte,  L. A. C. da Mota and  J. E. F. Skea,
{\em Solving second-order ordinary differential equations by
extending the Prelle-Singer method}. J. Phys. A  34  (2001),
3015--3024.

\bibitem{EE} N. Euler and M. Euler, {\em Sundman symmetries of nonlinear second-order and
third-order ordinary differential equations}.  J. Nonlinear Math.
Phys.  11  (2004),  no. 3, 399--421.

\bibitem{EESA} M. Euler, N. Euler, A. Str\"omberg and E. Str\"om, {\em Transformation between a
generalized Emden-Fowler equation and the first Painlev\'e
transcendent}. Math. Methods Appl. Sci.  30  (2007),  no. 16,
2121--2124.

\bibitem{EWLE} Euler, N.; Wolf, T.; Leach, P. G. L.; Euler, M. {\em
Linearisable third-order ordinary differential equations and
generalised Sundman transformations: the case $X'''=0$}.  Acta
Appl. Math.  76  (2003),  no. 1, 89--115.

\bibitem{GR} A Gallegos and H  C Rosu, {\it Comment Demystifying the constancy of the Ermakov-Lewis invariant 
for a time dependent oscillator}arXiv:1806.11139v1[math-ph],  Modern Phys. Lett. A 33 (2018), no. 24, 1875001, 3 pp.

\bibitem{GuGC} P. Guha and A. Ghose Choudhury, {\it Integrable Time-Dependent Dynamical Systems:
Generalized Ermakov-Pinney and Emden-Fowler Equations}, Nonlinear Dynamics and Systems Theory, 14 (4) (2014) 355-370.

\bibitem{GuGC2} P. Guha and A. Ghose Choudhury, {\it A note on generalization of the Ermakov-Lewis invariant and its demystification},
Mod. Phys. Lett. A34 (2019) 1950021.

\bibitem{GCK} P. Guha, A. Ghose Choudhury and B. Khanra, {\em On Solutions of Third and Fourth-Order
Time Dependent Riccati Equations and the Generalized Chazy
System}, Commun Nonlinear Sci Numer Simulat 17 (2012) 4053-4063.

%\bibitem{IR} R. Iacono and F. Russo, {\em Class of solvable nonlinear oscillators with isochronous
%orbits}, Phys. Rev. E 83 (2011) 027601.

\bibitem{Haas} F. Haas, {\em The damped Pinney equation and its applications to dissipative quantum mechanics}, Phys. Scr. 81 (2010) 025004 (7pp).

\bibitem{JS} D.W Jordan and P. Smith,  Nonlinear Ordinary
Differential Equations-{\em An introduction for Scientists and
Engineers} 4th Edition, Oxford University Press, 2007.

\bibitem{Leach} P.G.L. Leach and K. Andriopoulos, {\em The Ermakov Equation: a commentary},
Appl. Anal. Discrete Math. 2 (2008), 146-157.

\bibitem{Lie} S. Lie, {\em Klassifikation und integration von gewöhnlichen differentialgleichungen zwischen 
$x, y$, die eine gruppe von transformationen gestatten: III}, Arch. Mat. Naturvidenskab 8 (1883)  371-427 (Reprinted in Lie's Gessammelte Abhandlungen 1924 5 362-427, paper XIY)

\bibitem{Lienard} A. Li\'enard, A. (1928) {\em Etude des oscillations entretenues}, 
Revue g\'en\'erale de $l\sp{\prime}$ \'electricit\'e 23, pp. 901-912 and 946-954.

\bibitem{Liouville} R. Liouville, Sur les invariants de certaines e\'quations diffe\'rentielles et sur 
leurs applications, J. $l\sp{\prime}$ E\'ole Polytechnique 59 (1889), 7-76.

\bibitem{LV}) J. Llibre and C. Valls,
{\em Analytical integrability of the Rikitake system},
Z. Angew. Math. Phys. 61 (2010), no. 4, 627–634.
 
\bibitem{Meleshko} S. V. Meleshko, S. Moyo, C. Muriel, J. L. Romero, P. Guha, A. G. Choudhury, On
first integrals of second-order ordinary differential equations, Journal of Engineering
Mathematics October 2013, Volume 82, Issue 1, pp 1730.

\bibitem{Muriel} C. Muriel, J. L. Romero, Second-order ordinary differential equations and first
integrals of the form A(t, x) ˙x+B(t, x), Journal of Nonlinear Mathematical Physics,
Vol. 16, Suppl. (2009) 209222

\bibitem{Paddy} T Padmanabhan,{\it Demystifying the constancy of the Ermakov-Lewis invariant for a time
dependent oscillator}, Mod. Phys.
Letts A V 33, Nos. 7 \& 8 (2018) 1830005 ; arXiv:1712.07328v1[physics.class-ph] 20 Dec 2017.

\bibitem{RR1} J. R. Ray and J. L. Reid, {\em More exact invariants for the time dependent harmonic oscillator},
Phys Lett A 71 (1979) 317-318.

\bibitem{Sookmee}  S. Sookmee and S. V. Meleshko, {\em Conditions for linearization of a 
projectable system of two second-order ordinary differential equations}, J. Phys. A: Math. Theor. 41 402001.

\bibitem{Sund} K. F. Sundman, {\em M\'emoire sur le probl\'em des trois corps},
Acta Math. 36 (1912--1913), 105--179.

\bibitem{Saba1}M Sabatini, {\em On the period Function of
Li\'{e}nard Systems}, J. Diff. Eqns. 152,467-487, (1999).

\bibitem{SzB} Szebehely, V.; Bond, V., {\em  Transformations of the perturbed two-body problem
to unperturbed harmonic oscillators}, Celestial Mechanics and
Dynamical Astronomy Volume 30  (1983) 59-69.

\bibitem{Tresse} A. Tresse, D\'etermination des invariants ponctuels de l’ \'equation diff\'erentielle ordinaire du
second ordre $y^{\prime \prime} = ω(x, y, y^{\prime} )$, Leipzig, 1896.

\end{thebibliography}
\end{document}